\documentstyle[12pt]{article}

\title{Maple procedures in teaching the canonical formalism
of general relativity}

\author{Dumitru N. Vulcanov$^a$ and Gabriela Ciobanu$^b$\\
a) The West University of Timi\c soara\\
~~~Theoretical and Computational Physics Department\\
~~~V. P\^ arvan no. 4 Ave., 1900 Timi\c soara, Romania\\
and\\
b) ``Al.I Cuza'' University of Ia\c si\\
~~~Theoretical Physics Department\\
~~~Copou no. 11 Ave., 6600 Ia\c si, Romania}

\begin{document}
\date{}
\maketitle

\abstract{We present some Maple procedures using the GrTensorII package
for teaching purposes in the study of the canonical version of the general
relativity based on the ADM formalism}

\section{Introduction}

The use of computer facilities cam be an
important tool for teaching general relativity. We
have experienced several packages of procedures, (in REDUCE + EXCALC for
algebraic programming and in Mathematica for graphic visualizations) which
fulfill this purpose (\cite{10}). In this article we shall present some new
procedures in MapleV using GrTensorII package (\cite{11}) adapted for the 
canonical version of the general relativity (in the so called ADM formalism 
based on the 3+1 split of spacetime). This formalism is widely used 
(\cite{8},\cite{9}) in the last years as a major tool in numerical 
relativity for calculating violent processes as, for example the head-on 
collisions of black holes, massive stars or other astrophysical objects.
Thus we used these computer procedures in the process of teaching the canonical
formalism as an introductory part of a series of lectures on numerical
relativity for graduated students. 
The next section of the article presents shortly the notations and the main
features of the canonical version of the general relativity. 
Early attemps in using computer algebra (in REDUCE) for the ADM formalism
can be detected in the literature (\cite{3}, \cite{6},\cite{7}). Obviously
we used these programs in producing our new procedures for Maple + GrTensorII
package, but because there are many specific features we shall present in some
detail these procedures in the section 3 of the article. 
The last section of the article is dedicated to  the conclusions pointed
out 
by running the Maple procedures presented here and some future prospectives
on their usage toward the  numerical realativity.

\section{Review of the canonical formalism of general relativity}

 Here we shall use the specific notations for the ADM formalism 
\cite{1},\cite{2}; for example latin indices will run from 1  to  3  and  
greek 
indices  from  0  to  3.  The  starting  point  of  the  canonical 
formulation of the general  relativity  is  the  (3+1)-dimensional 
split of the space-time produced by the split of the metric tensor :
\begin{equation}\label{1}
   ^{(4)}g_{\alpha\beta} =   
         \left ( \begin{array}{ccc}       
           ^{(4)}g_{oo} & ~~ &  ^{(4)}g_{oj}       \\
                   ~~ & ~~ & ~~                    \\
           ^{(4)}g_{io} & ~~ &  ^{(4)}g_{ij}   \end{array} \right ) =
         \left ( \begin{array}{ccc}       
           N_{k}N^{k}-N^{2} & ~~ &   N_{j}        \\
                         ~~ & ~~ & ~~             \\
                 N_{i} & ~~ &   g_{ij}    \end{array} \right )
\end{equation}
    where $g_{ij}$ is the riemannian metric tensor of the three-dimensional 
spacelike hypersurfaces at t = const. which realize the  spacetime 
foliation.  Here $N$ is the  "lapse"  function and $N^{i}$ are the 
components of the "shift" vector \cite{2}.  

    The Einstein vacuum field equations now are (denoting by "$\cdot$"
the time derivatives) :  
\begin{equation}\label{2}
\dot{g}_{ij} = 2Ng^{-1/2}[\pi_{ij}-\frac{1}{2}g_{ij}{\pi^k}_k]+N_{i/j}+N_{j/i}
\end{equation}
\begin{eqnarray}
\dot{\pi}^{ij} = -Ng^{1/2}[R^{ij}-\frac{1}{2}g^{ij}R]+
       \frac{1}{2}Ng^{-1/2}g^{ij}[\pi^{kl}\pi_{kl}-\frac{1}{2}({\pi^k}_k)^{2}] 
\nonumber
\end{eqnarray}
\begin{eqnarray}
~~~~~~~~~~-2Ng^{-1/2}[\pi^{im}{\pi^j}_{m}-\frac{1}{2}\pi^{ij}{\pi^k}_k] +
                  g^{1/2}[N^{/ij}-g^{ij}{N^{/m}}_{/m}] \nonumber
\end{eqnarray}
\begin{equation} \label{3}
~~~~~~~~+[\pi^{ij}N^{m}]_{/m} - {N^i}_{/m}\pi^{mj} - {N^j}_{/m}\pi^{mi}
\end{equation}
where $\pi^{ij}$ are the components of the momenta canonically  conjugate 
to the $g_{ij}$'s.

    In the above formulas we denoted by "/" the  three-dimensional 
covariant derivative defined with $g_{ij}$ using the components of  the 
three-dimensional connection \cite{2} :
\begin{equation} \label{4}
\Gamma^{i}_{jk} = \frac{1}{2}g^{im}(g_{mj,k}+g_{mk,j}-g_{jk,m})
\end{equation}
The Ricci tensor components are given by
\begin{equation} \label{6}
  R_{ij}=\Gamma^{k}_{ij,k}-\Gamma^{k}_{ik,j}+\Gamma^{k}_{ij}\Gamma^{m}_{km}-
                 \Gamma^{k}_{im}\Gamma^{m}_{jk}
\end{equation}
    The initial data  on  the  t  =  const.  hypersurface  are  not 
independent because they must satisfy  the  constraint  equations, 
which complete the Einstein equations
\begin{equation} \label{7}
{\cal{H}}= -\sqrt{g} \lbrace R+g^{-1}[\frac{1}{2}({\pi^k}_{k})^{2}-\pi^{ij}
\pi_{ij}]\rbrace
                      =0
\end{equation}
\begin{equation} \label{8}
        {\cal{H}}^i=-2{\pi^{ij}}_{/j} =0
\end{equation}
where $\cal{H}$ is the super-hamiltonian, ${\cal{H}}^{i}$ the super-momentum 
and 
$g$ is the determinant of the three-dimensional metric tensor $g_{ij}$.

    The action functional in Hamiltonian form for a  vacuum  space 
-time can thus be written as (\cite{1},\cite{2}) :
\begin{equation} \label{9}
  S=\int{dt \int{ (\pi^{ij} \dot{g}_{ij} - N{\cal{H}} - N_{i}{\cal{H}}^{i})}}
                              \omega^{1}\omega^{2}\omega^{3}
\end{equation}
where the $\omega^i$'s are the basis one-forms.
    Thus  the  dynamic  equations  (\ref{2})  and  (\ref{3})
  are  obtained   by 
differentiating $S$ with respect to the canonical conjugate pair  of 
variables $(\pi^{ij},g_{km})$.

\section{Maple + GrTensorII procedures}

Here we shall describe briefly the structure and the main features of the Maple
procedures for the canonical formalism of the general relativity as described
in the previous section. Two major parts of the programs can be detected :
one before introducing the metric of the spacetime used (consisting in
several definitions of tensor objects which are common to all spacetimes)
and the second one,  having line-commands specific to each version.

The first part of the program starts after initalisation of the GrTensorII
package ({\bf grtw();}) and has mainly the next lines :
\begin{verbatim}
> grdef(`tr := pi{^i i}`);
> grdef(`ha0:=-sqrt(detg)*(Ricciscalar+
               (1/detg)*((1/2)*(tr)^2-pi{i j}*pi{ ^i ^j }))`);
> grdef(`ha{ ^i }:=-2*(pi{ ^i ^j ;j}-pi{ ^i ^j }*Chr{ p j ^p })`);
> grdef(`derge{ i j }:=2*N(x,t)*(detg)^(-1/2)*(pi{ i j } - 
                     (1/2)*g{ i j}*tr)+Ni{ i ;j } + Ni{ j  ;i }`);
> grdef(`Ndd{ ^m j }:= Nd{ ^m ;j }`); 
> grdef(`bum{ ^i ^j ^m}:=pi{ ^i ^j }*Ni{ ^m }`);
> grdef(`bla{ ^i ^j }:=bum{ ^i ^j ^m ;m }`);
> grdef(`derpi{ ^i ^j }:=
     -N(x,t)*(detg)^(1/2)*(R{ ^i ^j }-(1/2)*g{ ^i ^j }*Ricciscalar)+ 
      (1/2)*N(x,t)*(detg)^(-1/2)*g{ ^i ^j }*(pi{ ^k ^l   }*pi{ k l }-
      (1/2)*(tr)^2)-2*N(x,t)*(detg)^(-1/2)*(pi{ ^i ^m }*pi{ ^j m }-
      (1/2)*pi{ ^i ^j }*tr)+ (detg)^(1/2)*(Ndd{ ^i ^j }-g{ ^i ^j }*
       Ndd{ ^m m }) + bla{ ^i ^j } - Ni{ ^i ;m }*pi{ ^m ^j }-
       Ni{   ^j ;m }*pi{ ^m ^i }`);
\end{verbatim}
Here {\bf ha0} and {\bf ha\{~$\hat{}$ ~i ~\}} represents the superhamiltonian 
and the
supermomentum as defined in eqs. (\ref{7}) and (\ref{8}) respectively and
{\bf tr} is the trace of momentum tensor density $\pi^{ij}$ - which will
be defined in the next lines of the program. Here {\bf N(x,t)}represents the
lapse function $N$. Also, {\bf derge\{ ~i ~j~\}} 
represents the time derivatives of the components of the metric tensor,
as defined in eq. (\ref{2}) and {\bf derpi\{~$\hat{}$~i~~$\hat{}$ ~j~\}} the
time derivatives of the components of the momentum tensor $\pi^{ij}$ as
defined in eq. (\ref{3}). 

The next line of the program is a specific GrTensorII command for loading
the spacetime metric. Here Maple loads a file (previously generated) for
introducing the components of the metric tensor as functions of the 
coordinates.
We also reproduced here the output of the Maple session showing the metric 
structure of the spacetime we introduced. 
\begin{verbatim}
> qload(`Cyl_din`);

                     Default spacetime = Cyl_din
                      For the Cyl_din spacetime:
                             Coordinates
                                x(up)
                             a
                           x   = [x, y, z]
                             Line element
     2                                    2
   ds  = exp(gamma(x, t) - psi(x, t))  d x
           2                     2                       2
  + R(x, t)  exp(-psi(x, t))  d y   + exp(psi(x, t))  d z
\end{verbatim}
As is obvious we introduced above the metric for a spacetime with cylindrical
symmetry, an example we used for teaching purposes being a well known example
in the literature (\cite{5}). In natural output this metric has the form :
\begin{equation}\label{cyl}
g_{ij} = \left ( \begin{array}{ccc}
e^{\gamma-\psi}&0&0\\
0&R^2 e^{-\psi}&0\\
0&0&e^{\psi}\end{array}\right )
\end{equation}
in cylindrical coordinates $x,y,z$ with
           $x \in [0 , \infty)$, 
           $y \in [0 , 2\pi)$,
           $z \in (-\infty , +\infty)$
where $R$, $\psi$ and $\gamma$ are functions of $x$ and $t$ only.

After the metric of the spacetime is established the next sequence of the 
programm just introduce the components of the momentum tensor $\pi^{ij}$ as
\begin{verbatim}
> grdef(`Nd{ ^ m } := [diff(N(x,t),x), 0, 0]`);
> grdef(`Ni{ ^i } := [N1(x,t), N2(x,t), N3(x,t)]`);
> grdef(`vi1{^i}:=[pig(x,t)*exp(psi(x,t)-gamma(x,t)),0,0]`);
> grdef(`vi3{^i}
       :=[0,0,exp(-psi(x,t))*(pig(x,t)+(1/2)*R(x,t)*pir(x,t)+
                                      pip(x,t))]`);
> grdef(`vi2{^i}:=[0,(2*R(x,t))^(-1)*pir(x,t)*exp(psi(x,t)),0]`);
> grdef(`pi{ ^i ^j } := 
          vi1{ ^i }*kdelta{^j $x}+vi2{ ^i }*kdelta{ ^j$y }+
          vi3{ ^i }*kdelta{^j $z}`);
> grcalc(pi(up,up));
> grdisplay(pi(up,up));
\end{verbatim}
Here {\bf Ni\{~$\hat{}$~i~\}} represents the shift vector $N^i$ and the other
objects ({\bf Nd}, {\bf vi1}, {\bf vi2} and {\bf vi3}) represent intermediate
vectors defined in order to introduce the momenum {\bf pi\{ ~$\hat{}$~i~
$\hat{}$~j~\}} having the form :
\begin{equation}\label{mom}
\pi^{ij}= \left ( \begin{array}{ccc}
\pi_{\gamma} e^{\psi-\gamma}& 0 & 0\\
0 & \frac{1}{2R}\pi_R e^{\psi} & 0 \\
0 & 0 & e^{-\psi}(\pi_{\gamma}+\frac{1}{2}R\pi_R + \pi_{\psi})\end{array}
\right )
\end{equation}
In the program we denoted $\pi_{\gamma}$, $\pi_R$ and 
$\pi_{\psi}$ with {\bf pig}, {\bf pir} and {\bf pip}, respectively. 
 The momentum components are introduced in order  that the dynamic 
part of the action of the theory be in  canonical form, that is : $\dot{g}_{ij}
\pi^{ij} = \pi_{\gamma}\dot{\gamma} + \pi_{\psi}\dot{\psi}+\pi_R\dot{R}$.
The next lines of the programm check if this condition is fullfiled :
\begin{verbatim}
> grdef(`de1{ i }:=[diff(grcomponent(g(dn,dn),[x,x]),t),0,0]`);
> grdef(`de2{ i }:=[0,diff(grcomponent(g(dn,dn),[y,y]),t),0]`);
> grdef(`de3{ i }:=[0,0,diff(grcomponent(g(dn,dn),[z,z]),t)]`);
> grdef(`ddgt({ i j }:=
       de1{ i }*kdelta{j $x}+de2{ i }*kdelta{ j$y }+
       de3{ i }*kdelta{ j $z}`);
> grcalc(ddgt(dn,dn));
> grdef(`act:=pi{ ^i ^j }*ddgt{ i j }`);
> grcalc(act); gralter(act,simplify); grdisplay(act);
\end{verbatim}
By inspecting this last output from the Maple worksheet, the user can decide
if it is necessary to redifine the components of the momentum tensor or to go
further. Here the components of the momentum tensor were calculated by hand
but,
of course a more experienced user can try to introduce here a
sequence of commands for automatic calculation of the momentum tensor
components using the above condition, through an intensive use of {\bf solve}
Maple command.

Now comes the must important part of the routine, dedicated to calculations
of different objects previously defined :
\begin{verbatim}
> grcalc(ha0); gralter(ha0,simplify);
> grdisplay(ha0);
> grcalc(ha(up)); gralter(ha(up),simplify);
> grdisplay(ha(up));
> grcalc(derge(dn,dn)); gralter(derge(dn,dn),simplify);
> grdisplay(derge(dn,dn));
> d1:=exp(-psi(x,t))*grcomponent(derge(dn,dn),[z,z])+exp(psi(x,t)-
      gamma(x,t))*grcomponent(derge(dn,dn),[x,x]);
> simplify(d1);
> d2:=(1/(2*R(x,t)))*exp(psi(x,t))*grcomponent(derge(dn,dn),[y,y])+
       (1/2)*R(x,t)*exp(-psi(x,t))*grcomponent(derge(dn,dn),[z,z]);
> simplify(d2);
> d3:=exp(-psi(x,t))*grcomponent(derge(dn,dn),[z,z]);
> simplify(d3);
> grcalc(derpi(up,up)); gralter(derpi(up,up),simplify);
> grdisplay(derpi(up,up));
> f1 := exp(gamma(x,t)-psi(x,t))*grcomponent(derpi(up,up),[x,x])-
                                                pig(x,t)*(d3-d1);
> simplify(f1);
>  f2:= 2*R(x,t)*exp(-psi(x,t))*grcomponent(derpi(up,up),[y,y])+
                             (1/R(x,t))*d2*pir(x,t)-pir(x,t)*d3;
> simplify(f2);
> f3 := exp(psi(x,t))*grcomponent(derpi(up,up),[z,z])+d3*(pig(x,t)+
        (1/2)*R(x,t)*pir(x,t)+pip(x,t))-f1-(1/2)*R(x,t)*f2-
                                                 (1/2)*pir(x,t)*d2;
> simplify(f3);
\end{verbatim}
This is a simple series of alternation of{\bf grcalc}, {\bf gralter} and
{\bf grdisplay} commands for obtainig the superhamiltonian, supermomentum
and the dynamic equations for the theory. {\bf d1 ... d3} and {\bf f1 ... f3}
are the time derivatives of the dynamic variables, $\dot{\gamma}$, $\dot{R}$,
$\dot{\psi}$ and $\dot{\pi}_{\gamma}$, $\dot{\pi}_{R}$, $\dot{\pi}_{\psi}$
respectively.
 Denoting with "$\prime$" the derivatives with respect to $r$ we display here
the results for the example used above (cylindrical gravitational waves)  :
\begin{eqnarray} 
    {\cal{H}}^0=e^{\frac{\psi-\gamma}{2}}
                  ( 2R^{\prime\prime} - R^{\prime}\gamma^{\prime} + 
                    \frac{1}{2}(\psi^{\prime})^{2}R -\pi_{\gamma}\pi_{R} +
                 \frac{1}{2R}(\pi_{\psi})^2 ) = 0 \hbox{~~~~~~~~~~~~~~~~~~} 
\nonumber
\end{eqnarray}
\begin{eqnarray} 
    {\cal{H}}^1={\cal{H}}^r=e^{\psi-\gamma}
                  ( -2\pi^{\prime}_{\gamma}+\gamma^{\prime}\pi_{\gamma}+
                    R^{\prime}\pi_R + \psi^{\prime}\pi_{\psi} ) = 0
                  \hbox{~~~~~~;~~~~~~}   {\cal{H}}^2 = {\cal{H}}^3 = 0 
\nonumber
\end{eqnarray}
\begin{eqnarray} 
    \dot{\gamma} = N^1\gamma^{\prime} +2N^{1\prime} - 
                               e^{ \frac{\psi-\gamma}{2}}N\pi_{R} 
\hbox{~~~~~;~~~~~}
    \dot{R} = N^1R^{\prime} - e^{\frac{\psi-\gamma}{2}}N\pi_{\gamma} 
\hbox{~~~~~~}
\nonumber
\end{eqnarray}
\begin{eqnarray} 
      \dot{\psi} = N^1\psi^{\prime} + \frac{1}{R}e^{\frac{\psi-\gamma}{2}}
N\pi_{\psi}
\hbox{~~~~~~;~~~~~~} 
\nonumber
\end{eqnarray}
\begin{eqnarray} 
     \dot{\pi}_{\gamma} = N^1 \pi_{\gamma}^{\prime} + N^{1 \prime} \pi_{\gamma}
                    -e^{ \frac{\psi - \gamma}{2} } ( R^{\prime} N^{\prime}
                 + \frac{1}{2} R^{\prime} \psi^{\prime} N 
                      - \frac{1}{4} \psi^{\prime 2} R N 
 + \frac{1}{2} N \pi_{\gamma}  \pi_R 
                - \frac{1}{4R} N \pi_{\psi}^2 )
\nonumber
\end{eqnarray}
\begin{eqnarray} 
      \dot{\pi}_{R} =  N^1 \pi_R^{\prime} + N^{1 \prime} \pi_R +
                   e^{\frac{\psi-\gamma}{2}} ( \gamma^{\prime} N^{\prime} 
             -2 N^{\prime \prime} - 2 N^{\prime} \psi^{\prime} 
                +\frac{1}{2} \gamma^{\prime} \psi^{\prime} N 
        - \psi^{\prime \prime} N - \psi^{\prime 2}
             +\frac{1}{2R} N \pi_{\psi}^2 ) 
\nonumber
\end{eqnarray}
\begin{eqnarray} 
         \dot{\pi}_{\psi} = N^1 \pi_{\psi}^{\prime} + N^{1 \prime} \pi_{\psi}
            + e^{ \frac{\psi -\gamma}{2} } ( R N^{\prime} \psi^{\prime}
             - R^{\prime \prime} N + \frac{1}{2} N R^{\prime} \gamma^{\prime}
                  + R^{\prime} \psi^{\prime} N 
             -\frac{1}{2} \gamma^{\prime} \psi^{\prime} N R
\nonumber
\end{eqnarray}
\begin{eqnarray} 
              +  \psi^{\prime \prime} R N + \frac{1}{4} \psi^{\prime 2} R N
              + \frac{1}{2} N \pi_R \pi_{\gamma} - \frac{1}{4R} N 
\pi_{\psi}^2 ) 
\nonumber
\end{eqnarray}
These are the well-known results reported in (\cite{5}) or (\cite{6}).

One of the important goals of the canonical formalism of the general relativity
(which constitutes the ``kernel'' of the ADM formalism) is the reductional
formalism. Here we obtain the true dynamical status of the theory, by reducing 
the number of the variables  through solving the constraint
equations. This formalism is applicable only to a restricted number of
space-time models, one of them being the above cylindrical gravitational
waves model. Unfortunately only a specific strategy can be used in every
model. Thus the next lines of our program must be rewritten specifically in
every case. Here, for teaching purposes we present our example of 
cylindrical gravitational wave space-time model. Of course we enccourage the 
student to apply his own strategy for other examples he dares to calculate.

In our example of cylindrical gravitational waves, the reductional
strategy as described in (\cite{5}) starts with the usual rescaling 
 of ${\cal{H}}$ and ${\cal{H}}^i$ to $\bar{{\cal{H}}}$ 
and $\bar{{\cal{H}}}^i$ by
\begin{eqnarray} 
  \bar{{\cal{H}}} = e^{ \frac{\gamma-\psi}{2} }{\cal{H}} \hbox{~~;~~}
          \bar{N} = e^{ \frac{\psi-\gamma}{2} }N
               \hbox{~~;~~}
  \bar{{\cal{H}}}^1 = e^{ \gamma-\psi }{\cal{H}}^1 \hbox{~~;~~}
          \bar{N}^1 = e^{ \psi-\gamma }N^1
\nonumber
\end{eqnarray}  
wich produce the next sequence of Maple+GrTensorII commands :
\begin{verbatim}
> grdef(`aha0:=sqrt(exp(gamma(x,t)-psi(x,t)))*ha0`);
> grdef(`aha{ ^j } := exp(gamma(x,t)-psi(x,t))*ha{ ^j }`);
> grdef(`an:=sqrt(exp(psi(x,t)-gamma(x,t)))*n(x,t)`);
> grdef(`ani{ ^i } := exp(psi(x,t)-gamma(x,t))*ni{ ^i }`);
\end{verbatim}
The canonical transformation to the new 
variables, including Kuchar's "extrinsic time", defined by :
\begin{eqnarray} 
    T = T(\infty) + \int_{\infty}^{r}(-\pi_{\gamma})dr
                \hbox{~~~,~~~}
    \Pi_{T} = -\gamma^{\prime} + [\ln{((R^{\prime})^2-(T^{\prime})^2})
]^{\prime}
\nonumber
\end{eqnarray}
\begin{eqnarray} 
    R = R
            \hbox{~~~,~~~}
    \Pi_{R} = \pi_{R} + 
       [\ln{(\frac{R^{\prime}+T^{\prime}}{R^{\prime}-T^{\prime}})}]^{\prime}
\nonumber
\end{eqnarray}
are introduced with :
\begin{verbatim}
> pig(x,t):=-diff(T(x,t),x);

> pir(x,t):=piR(x,t) - diff(ln((diff(R(x,t),x)+diff(T(x,t),x))/
                           (diff(R(x,t),x)-diff(T(x,t),x))),x);
\end{verbatim}
and specific substitutions in the dynamic objects of the theory :
\begin{verbatim}
> grmap(ha0, subs , diff(gamma(x,t),x)=
diff( ln( (diff(R(x,t),x))^2- (diff(T(x,t),x))^2 ),x)-piT(x,t),`x`);
> grcalc(ha0); gralter(ha0,simplify);
> grdisplay(ha0);
> grmap(ha(up), subs , diff(gamma(x,t),x)=diff( ln( (diff(R(x,t),x))^2- 
          (diff(T(x,t),x))^2 ),x)-piT(x,t),`x`);
> gralter(ha(up),simplify);
> grdisplay(ha(up));
> grcalc(aha0);
> grmap(aha0, subs , diff(gamma(x,t),x)=diff( ln( (diff(R(x,t),x))^2- 
                  (diff(T(x,t),x))^2 ),x)-piT(x,t),`x`);
> gralter(aha0,simplify,sqrt);
> grdisplay(aha0);
> grcalc(aha(up));
> grmap(aha(up), subs , diff(gamma(x,t),x)=diff( ln( (diff(R(x,t),x))^2- 
                  (diff(T(x,t),x))^2 ),x)-piT(x,t),`x`);
> gralter(aha(up),simplify);
> grdisplay(aha(up));
> grmap(act, subs , diff(gamma(x,t),x)=diff( ln( (diff(R(x,t),x))^2- 
                (diff(T(x,t),x))^2 ),x)-piT(x,t),`x`);
> grcalc(act); grdisplay(act);
\end{verbatim}
     Thus the action yields (modulo divergences) :
\begin{eqnarray} 
   S = 2 \pi \int_{-\infty}^{\infty} dt \int_{0}^{\infty} dr 
                  (\Pi_T \dot{T} + \Pi_R \dot{R} + \pi_{\psi} \dot{\psi}
                             + \pi_{\chi} \dot{\chi} -
                  \bar{N} \bar{{\cal{H}}} - \bar{N}_1 \bar{{\cal{H}}}^1 )
\nonumber
\end{eqnarray}
    where :
\begin{eqnarray} 
   \bar{{\cal{H}}} = R^{\prime} \Pi_T + T^{\prime} \Pi_R +
                \frac{1}{2}R^{-1} \pi_{\psi}^2 + \frac{1}{2}R \psi^{\prime 2}
                  +\frac{1}{4}R^{-1} \pi_{\chi}^2 + R \chi^{\prime 2}
\nonumber
\end{eqnarray}
\begin{eqnarray} 
   \bar{{\cal{H}}}^1 = T^{\prime} \Pi_T + R^{\prime} \Pi_R + 
                  \psi^{\prime} \pi_{\psi} +
                        \chi^{\prime} \pi_{\chi}
\nonumber
\end{eqnarray}
 
Solving the constraint equations  $\bar{{\cal{H}}}=0$ and 
$\bar{{\cal{H}}}^1=0$ for $\Pi_T$ and $\Pi_R$ and
imposing the coordinate conditions $T = t$ and $R = r$ we obtain finally :
\begin{eqnarray} 
    S = 2 \pi \int_{-\infty}^{+\infty} dT \int_{0}^{+\infty} dR 
              [\pi_{\psi} \psi_{,T} + \pi_{\chi} \chi_{,T} - 
                \frac{1}{2}( R^{-1} \pi_{\psi}^2 + R \psi_{,R}^2
                  +  R \pi_{\chi}^2 + R^{-1} \chi^{\prime 2}  )]
\nonumber
\end{eqnarray}
from the next sequence of programm lines :
\begin{verbatim}
> R(x,t):=x; T(x,t):=t; grdisplay(aha0);
> solve(grcomponent(aha0),piT(x,t));
> piT(x,t):= -1/2*(x^2*diff(psi(x,t),x)^2+pip(x,t)^2)/x;
> eval(piR(x,t));
> piR(x,t):=-diff(psi(x,t),x)*pip(x,t); piR(x,t);
> grdisplay(aha0); grdisplay(aha(up));
> piT(x,t);

                       2 /d           \2            2
                      x  |-- psi(x, t)|  + pip(x, t)
                         \dx          /
                - 1/2 -------------------------------
                                     x

> piR(x,t);

                       /d           \
                      -|-- psi(x, t)| pip(x, t)
                       \dx          /

> grcalc(act); grdisplay(act);
                      For the Cyl_din spacetime:

                                 act

                          /d           \
                    act = |-- psi(x, t)| pip(x, t)
                          \dt          /

> grdef(`Action:=act+piT(x,t)*diff(T(x,t),t)+piR(x,t)*diff(R(x,t),t)`);
> grcalc(Action);gralter(Action,factor,normal,sort,expand);
> grdisplay(Action);
                      For the Cyl_din spacetime:

                                Action


                   /d           \2   /d           \
  Action = - 1/2 x |-- psi(x, t)|  + |-- psi(x, t)| pip(x, t)
                   \dx          /    \dt          /

                        2
               pip(x, t)
         - 1/2 ----------
                   x

> grdef(`Ham:=piT(x,t)*diff(T(x,t),t)+piR(x,t)*diff(R(x,t),t)`);
> grcalc(Ham); gralter(Ham,expand);
> grdisplay(Ham);
                      For the Cyl_din spacetime:

                                 Ham
                                                         2
                          /d           \2       pip(x, t)
            Ham = - 1/2 x |-- psi(x, t)|  - 1/2 ----------
                          \dx          /            x

\end{verbatim}

\section{Conclusions. Further improuvements}

We used the programms presented above in the computer room with the students
from the graduate course on Numerical Relativity. The main purpose was to
introduce faster the elements of the canonical version of relativity with
the declared objective to skip the long and not very straitforward  hand 
calculations necessary to process an entire example of spacetime model. We
encouraged the students to try to modify the procedures in order to
compute new examples. 

The major conclusion is that this method is indeed usefull for an attractive
and fast teaching of the methods involved in the ADM formalism. On the
other hand we can use and modify these programs for obtaining the equations
necessary for the numerical relativity. In fact we intend to expand our
Maple worksheets for the case of axisymmetric model (used in the numerical
treatement of the head-on collision of black-holes). Of course, for numerical
solving of the dynamic equations obtained here we need more improuvements of
the codes for paralel computing and more sophisticated numerical methods.
But this will be the object of another series of articles.

\end{document}